\documentclass[aps,pra,twocolumn,groupedaddress,showpacs,showkeys]{revtex4}

\usepackage{graphicx}
\usepackage{bm}
\usepackage{amsmath}

\begin{document}

\title{Microwave potentials and optimal control for robust quantum gates on an atom chip}

\author{Philipp~Treutlein}\email[E-mail: ]{philipp.treutlein@physik.lmu.de}
\author{Theodor~W.~H{\"a}nsch}
\author{Jakob~Reichel$^\ddag$}
\affiliation{Max-Planck-Institut f{\"u}r Quantenoptik und Sektion Physik der Ludwig-Maximilians-Universit{\"a}t, 80799 M{\"u}nchen, Germany\\
$^\ddag$Present address: Laboratoire Kastler Brossel de l'E.N.S, Paris, France}
\author{Antonio~Negretti}\email[E-mail: ]{negretti@phys.au.dk }
\altaffiliation[Also at: ]{Institut f\"ur Physik, Universit\"at Potsdam, Am Neuen Palais 10, 14469 Potsdam, Germany\\and Danish National Research Foundation Center for Quantum Optics, Department of Physics and Astronomy, University of {\AA}rhus, 8000 {\AA}rhus C, Denmark}
\author{Markus~A.~Cirone}
\author{Tommaso~Calarco$^\S$}
\affiliation{Dipartimento di Fisica, Universit\`a di Trento, and CNR-INFM-BEC, 38050 Povo (TN), Italy,\\ and ECT*, Strada delle Tabarelle 286, 38050 Villazzano (TN), Italy\\
$^\S$Present address: ITAMP, Harvard University, Cambridge, MA 02138, U.S.A.}

\date{\today}

\begin{abstract}
We propose a two-qubit collisional phase gate that can be implemented with available atom chip technology, and present a detailed theoretical analysis of its performance.
The gate is based on earlier phase gate schemes, but uses a qubit state pair with an experimentally demonstrated, very long coherence lifetime.
Microwave near-fields play a key role in our implementation as a means to realize the state-dependent potentials required for conditional dynamics. Quantum control algorithms are used to optimize gate performance.
We employ circuit configurations that can be built
with current fabrication processes, and extensively discuss the
impact of technical noise and imperfections that characterize an
actual atom chip. We find an overall infidelity compatible with
requirements for fault-tolerant quantum computation.
\end{abstract}


\maketitle

\section{Introduction}

The physical implementation of scalable quantum computing unquestionably poses huge challenges. Many proposals exist, most of which in principle satisfy all requirements for scalability \cite{diVincenzo00}. One might be tempted to claim that their realization is simply a matter of technological improvement. Yet, we can fairly say that to date nobody knows how to build a scalable quantum computer. Understanding and overcoming each of the sources of imperfection, specific to a given physical scenario, becomes thus crucial on the way to a possible actual implementation.

Atom chips \cite{Reichel02,Folman02} combine many important features of a scalable architecture for quantum information processing (QIP) \cite{diVincenzo00}: The long coherence lifetimes of qubits based on hyperfine states of neutral atoms \cite{Treutlein04}, accurate control of the coherent evolution of the atoms in tailored micropotentials \cite{Hommelhoff05,Schumm05}, scalability of the technology through microfabrication \cite{Lev03,Groth04}, which allows the integration of many qubits in parallel on the same device while maintaining individual addressability, and the exciting perspective of interfacing quantum optical qubits with solid-state systems for QIP located on the chip surface \cite{Sorensen04}. However, the experimental demonstration of a fundamental two-qubit quantum gate on an atom chip is an important milestone which still has to be reached.

A first proposal for a quantum phase gate based on collisional interactions between atoms on a chip was put forward in \cite{Calarco00}. While it demonstrates the working principle of such a gate, there are problems which could prevent a successful experimental realization: 1) The qubit is encoded in two states with a magnetic-field sensitive energy difference, such that it is hard to maintain the qubit coherence over a long time in a noisy experimental environment. 2) The fidelity is strongly reduced by wave packet distortion due to undesired collisions in some of the qubit basis states. 3) An idealized situation with accurately harmonic potentials for the atoms was considered \cite{Calarco00}, which is hard to realize experimentally, and deviations from harmonicity spoil the gate performance \cite{Negretti03}. 4) Furthermore, transverse excitations of the atoms were not considered in detail. For a successful experimental implementation, however, a scheme is needed which allows for high fidelity gate operations under realistic conditions. A gate infidelity (error rate) below a certain threshold is needed in order to allow for a fault-tolerant implementation of quantum computing. Depending on error models and recovery schemes, estimates of such threshold vary from a few $10^{-3}$ for active error correcting codes \cite{Steane03} up to well above $10\%$ for error detection schemes \cite{Knill05}.

In this paper, we present a substantially improved version of the phase gate of \cite{Calarco00} for a qubit state pair with an experimentally demonstrated coherence lifetime exceeding 1\,s \cite{Treutlein04}, and give a detailed prescription for its implementation. The decoherence rates of the qubits due to magnetic-field noise are suppressed by a factor of $10^{-3}$ compared to \cite{Calarco00}. In our proposal, a key role is played by microwave near-field potentials on the atom chip \cite{Treutlein04}, which allow to create the required state-selective potentials for a successful gate operation with the states considered here. Using microwave potentials, the same robust qubit states can be used for information processing and storage.  At the same time, the microwave potentials avoid the unwanted collisions limiting the fidelity in the original proposal, by appropriately displacing the potential minima for qubit states $|0\rangle$ and $|1\rangle$. We simulate the gate dynamics in a potential created by a realistic atom chip, which we specify in detail, and which can be fabricated with today's technology. Furthermore, we consider many sources of infidelity in detail, such as loss and decoherence effects due to the proximity of the chip surface, and we find a total infidelity of the order of a few $10^{-3}$.


\section{The collisional phase gate}

\subsection{Qubit states and single qubit rotations}

Quantum information processing with high fidelity is only possible if qubit basis states are chosen whose energy difference is sufficiently robust against noise in realistic experimental situations.
In particular, technical fluctuations of magnetic fields are notorious for limiting the coherence lifetime of magnetic-field sensitive qubit states of atoms or ions to a few milliseconds \cite{Schmidt-Kaler03}. On atom chips, magnetic near-field noise due to thermally excited currents in the chip wires is an additional fundamental source of decoherence for magnetic field sensitive qubit states \cite{Henkel03}. To achieve long coherence lifetimes on atom chips, it is therefore highly desirable to choose a pair of qubit basis states whose energy difference is insensitive to magnetic field fluctuations.
We choose the $|F=1,m_F=-1\rangle \equiv |0\rangle$ and
$|F=2,m_F=+1\rangle \equiv |1\rangle$ hyperfine levels of the $5S_{1/2}$ ground state of $^{87}$Rb atoms as qubit basis states. The magnetic moments and the corresponding static Zeeman shifts of the two states are approximately equal, leading to a strong common mode suppression of magnetic field induced decoherence. Furthermore, both states experience nearly identical trapping potentials in magnetic traps, thereby avoiding undesired entanglement between internal and external degrees of freedom of the atoms. Coherence lifetimes exceeding 1\,s have been achieved for this state pair with atoms in a magnetic trap at a distance of a few microns from the chip surface \cite{Treutlein04}. Single-qubit rotations can be easily implemented by coupling the states $|0\rangle$ and $|1\rangle$ through a two-photon microwave-rf transition, as demonstrated in \cite{Treutlein04}.

\subsection{Principle of the gate operation}

To complete the toolbox for quantum information processing on atom chips \cite{Treutlein06}, a fundamental two-qubit quantum gate is needed. Each qubit is represented by an atom in a superposition of the robust qubit states $|0\rangle$ and $|1\rangle$.
Our goal is to realize a phase gate, with the truth table
\begin{eqnarray}
|00\rangle &\rightarrow& |00\rangle, \nonumber \\
|01\rangle &\rightarrow& |01\rangle, \nonumber \\
|10\rangle &\rightarrow& |10\rangle, \nonumber \\
|11\rangle &\rightarrow& e^{i\phi_g} |11\rangle, \label{eq:TruthTable}
\end{eqnarray}
for the four two-qubit basis states. $\phi_g$ is the gate phase, which is to be adjusted to $\phi_g=\pi$. As proposed in \cite{Calarco00}, the phase gate can be implemented by modulating the trapping potential state-selectively, such that the two atoms interact and pick up a collisional phase shift $\phi_g$ if and only if both are in internal state $|1\rangle$. In the following, we briefly sketch the working principle of the gate, highlighting the differences between the present approach and \cite{Calarco00}.

\begin{figure}[]
  \includegraphics[scale=0.6]{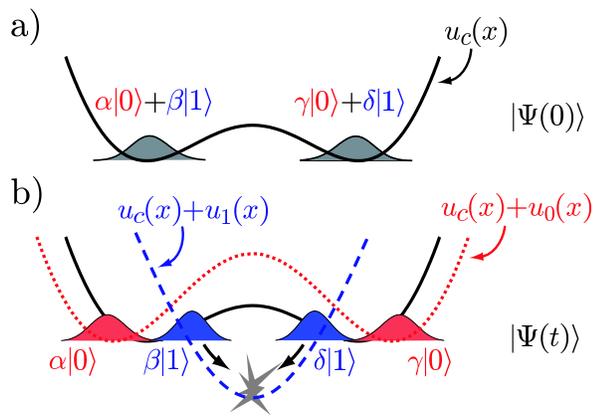}
\caption{\label{fig:StateSelPot} (Color online) State-selective potential, atomic wave functions, and principle of the gate operation. (a) The state-independent potential $u_c(x)$ along $x$ for $t<0$ and $t>\tau_g$, before and after the gate operation, when $\lambda(t)=0$. The initial state wave function of the two atoms in this potential is shown. (b) The state-dependent potential $u_c(x)+u_i(x)$ (here $\lambda(t)=1$) for $0\leq t \leq \tau_g$, during the gate operation. The atomic wave functions after half an oscillation period are shown. The state-independent part $u_c(x)$ is shown for comparison.
}
\end{figure}

In Figure~\ref{fig:StateSelPot}, the principle of the gate operation is shown. The atoms are placed in a state-dependent potential
\begin{equation}\label{eq:StateSelPot}
U_i(\mathbf{r},t) = u_c(\mathbf{r}) + \lambda(t)\cdot u_i(\mathbf{r}),
\end{equation}
which can be split into a common part $u_c(\mathbf{r})$ and a state-dependent part $u_i(\mathbf{r})$, where $i=\{0,1\}$ denotes the states $|0\rangle$ and $|1\rangle$ and $\mathbf{r}=(x,y,z)$. The common part of the potential is time-independent, while the state-dependent part is modulated with a function $\lambda(t)$, $0 \leq \lambda(t) \leq 1$, during the gate operation. At times $t<0$, when the gate is in its initial state, we have $\lambda(t) = 0$ and the atoms are subject to $u_c(\mathbf{r})$ only, see Fig.~\ref{fig:StateSelPot}a. The potential $u_c(\mathbf{r})$ provides a tight confinement in the transverse dimensions $y$ and $z$, such that the dynamics of the atoms is effectively one-dimensional. In the longitudinal dimension $x$, $u_c(\mathbf{r})$ is a double well with a sufficiently high barrier to prevent tunneling between the wells. Each of the qubit atoms is prepared in the motional ground state of one potential well. The gate operates during the time $0\leq t \leq \tau_g$, where $\tau_g$ is the gate time. During this time, $\lambda(t) \neq 0$, and the potential is state-dependent as sketched in Fig.~\ref{fig:StateSelPot}b. The effect of $u_i(\mathbf{r})$ is twofold: $u_1(\mathbf{r})$ removes the barrier of the double well for state $|1\rangle$, leaving only a single, approximately (but not exactly) harmonic potential well in which atoms in state $|1\rangle$ start to oscillate. $u_0(\mathbf{r})$ shifts the minima of the double well for state $|0\rangle$ further apart in the $x$-direction. The effect of $u_i(\mathbf{r})$ on the tight transverse confinement is very small. In this way, the truth table (\ref{eq:TruthTable}) is implemented: In state $|11\rangle$, both atoms will oscillate and collide each time they pass the center of the trap, which leads to the desired collisional phase shift of the state $|11\rangle$. In states $|00\rangle$, $|01\rangle$, and $|10\rangle$, the atoms do not collide, since atoms in state $|0\rangle$ are shifted out of the way of the oscillating state $|1\rangle$. When the desired phase shift $\phi_g$ is accumulated after an integer number of oscillations $N$ of the state $|1\rangle$, the gate operation is terminated by returning to $\lambda(t)=0$ for $t>\tau_g$, recapturing each atom in one of the potential wells of $u_c(\mathbf{r})$ (see again Fig.~\ref{fig:StateSelPot}a).

The gate operation described here is different from the original proposal in \cite{Calarco00}. There, only the potential for state $|1\rangle$ is switched during the gate operation. This leads to unwanted collisions of atoms in state $|01\rangle$, which are a major source of infidelity and would require an additional transverse shift in the potential minimum \cite{Calarco00}. Using microwave potentials, we are able to create a state-dependent potential in which the state $|0\rangle$ is shifted out of the way of the oscillating state $|1\rangle$ in the longitudinal dimension. This is essential to avoid the unwanted collisions and to achieve high fidelity gate performance. A second difference is that we choose a smooth modulating function $\lambda(t)$ for the state-selective part of the potential instead of instantaneous switching as in \cite{Calarco00}. Roughly speaking, $\lambda(t)$ is ramped from $\lambda(t)=0$ to $\lambda(t)=1$ during the first half of each oscillation, while it is ramped back to $\lambda(t)=0$ with the inverse temporal profile during the second half. This smooth ramping allows for much better control over the gate dynamics than instantaneous switching of the potential. This is necessary to avoid strong excitations of the atoms in $|0\rangle$ during the shift of the potential well. It also decreases the oscillation amplitude of state $|1\rangle$, thereby further suppressing collisions in the states $|01\rangle$ and $|10\rangle$. The exact time dependence of the function $\lambda(t)$ is determined by an optimal control algorithm which optimizes the fidelity of the quantum gate, as discussed in Section~\ref{sec:Optimization}.


\section{Microwave potentials}

The state-dependent potential (\ref{eq:StateSelPot}) which is needed for the gate operation can be realized by a combination of static magnetic and microwave fields on the atom chip, as briefly described in \cite{Treutlein04}.
The concept of microwave potentials is similar to the optical potentials created by non-resonant laser beams, which have been used with enormous success to generate potentials for the manipulation of ultracold atoms \cite{Bloch05}. The main difference is that spontaneous emission is negligible for microwave transitions, so that one can use (near) resonant coupling.

In $^{87}$Rb, microwave potentials derive from magnetic dipole transitions with a frequency near $\omega_0/2\pi= 6.835$\,GHz between the $F=1$ and $F=2$
hyperfine manifolds of the ground state (Fig.~\ref{fig:HFSmw}). The magnetic component of the microwave field couples the $|F=1,m_1\rangle$ to the $|F=2,m_2\rangle$ sublevels and leads to energy shifts that depend on $m_1$ and $m_2$. In a spatially varying microwave field, this results in a state-dependent potential landscape. In addition, there is a small common mode part of the potential due to the electric field of the microwave, which shifts all levels $|F,m_F\rangle$ identically.

A trap for neutral atoms based on microwave potentials has been proposed in \cite{Agosta89} and experimentally demonstrated in \cite{Spreeuw94}. This trap employs microwave radiation in the far field of the source. Unlike the case of optical radiation, which can be tightly focussed due to its short wavelength, the centimeter wavelength $\lambda_\textrm{mw}$ of microwave radiation poses severe limitations on far-field traps: field gradients are very weak \cite{Spreeuw94} and structuring the potential on the micrometer scale is impossible.

On atom chips, there is a natural solution to this problem. The atoms are trapped at distances $d\ll \lambda_\textrm{mw}$ from the chip surface.  Thus, they can be manipulated with microwave near fields, generated by microwave signals in on-chip transmission lines \cite{Collin01}. In the near field of the source currents and voltages, the microwave fields have the same position dependence as the static fields created by equivalent stationary sources. The maximum field gradients depend on the size of the transmission line conductors and on the distance $d$, not on $\lambda_\textrm{mw}$. In this way, state-dependent microwave potentials varying on the micrometer scale can be realized. In combination with state-independent static magnetic microtraps, the complex potential geometries required for QIP can be realized.

The use of microwaves to create the state-selective potential is essential for our proposal, since a combination of static magnetic and electric fields, as considered in \cite{Calarco00,Krueger03}, does not provide state-selective potentials for our robust qubit state pair, whose magnetic moments and electrostatic polarizabilities are equal.
Optical potentials created by focussed laser beams with a frequency close to the D1 or D2 transition of $^{87}$Rb are also impractical: if the detuning of the laser from the atomic resonance is much larger than the hyperfine splitting of the $^{87}$Rb ground state, the resulting optical potentials are again nearly identical for the states $|0\rangle$ and $|1\rangle$. If, on the other hand, a detuning comparable to the hyperfine splitting is used, a differential optical potential could be created, but problems with decoherence due to spontaneous scattering of photons arise.

\subsection{Theory of microwave potentials \label{ssec:MWpot}}

We consider the hyperfine levels $|F,m_F\rangle$ of the $5S_{1/2}$ ground state of a $^{87}$Rb atom in a static magnetic field $\mathbf{B}_0(\mathbf{r})$. In addition, the atom is exposed to a microwave of frequency $\omega$ with magnetic field $\mathbf{B}_\mathrm{mw}(\mathbf{r})\cos(\omega t)$ and electric field $\mathbf{E}_\mathrm{mw}(\mathbf{r})\cos(\omega t + \varphi)$, where $\varphi$ accounts for a phase shift between electric and magnetic field. We now calculate the adiabatic trapping potentials for the atomic hyperfine levels in the combined static and microwave fields. The trapping frequencies considered in this paper are sufficiently small so that non-adiabatic transitions due to the atomic motion do not limit the trap lifetime \cite{Folman02}.

\begin{figure}[]
  \includegraphics[scale=0.6]{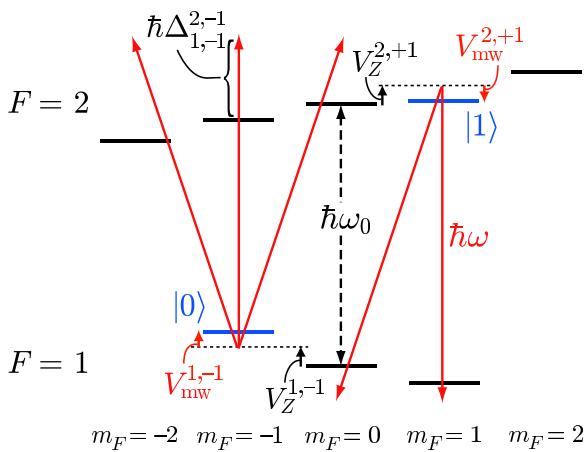}
\caption{\label{fig:HFSmw} (Color online) Energy-level diagram of the hyperfine structure of the $^{87}$Rb ground state in a combined static magnetic and microwave field. $V_Z^{F,m_F}$ indicates the energy shift due to the static Zeeman effect, which is (nearly) identical for $|0\rangle$ and $|1\rangle$. Due to the magnetic field of the microwave, the level $|F,m_F\rangle$ is shifted in energy by $V_\mathrm{mw}^{F,m_F}$. This shift has opposite sign for $|0\rangle$ and $|1\rangle$, as indicated in the figure (the shift of the other levels is not shown). The microwave transitions contributing to the shift of $|0\rangle$ and $|1\rangle$ are shown for $\Delta_{1,m_1}^{2,m_2} > 0$ (blue detuning).
}
\end{figure}

The static magnetic field $\mathbf{B}_0(\mathbf{r})$ leads to a Zeeman shift
\begin{equation} \label{eq:VZ}
V_Z^{F,m_F} (\mathbf{r}) = \mu_B g_F m_F |\mathbf{B}_0(\mathbf{r})|
\end{equation}
of the levels (see Fig.~\ref{fig:HFSmw}), where $\mu_B$ is the Bohr magneton and $g_F$ is the hyperfine Land\'e factor for the state $F$. This shift is identical for states $|0\rangle$ and $|1\rangle$ and we identify it with the common potential in (\ref{eq:StateSelPot}),
\begin{equation}
u_c (\mathbf{r}) = V_Z^{2,+1} (\mathbf{r}) = V_Z^{1,-1} (\mathbf{r}) = \frac{\mu_B}{2}|\mathbf{B}_0(\mathbf{r})|.
\end{equation}
Effects of the nuclear magnetic moment have been neglected in (\ref{eq:VZ}), they lead to corrections in the potential of the order of $10^{-3}$. Here and in the following, we take the local direction of the static field $\mathbf{B}_0(\mathbf{r})$ as the quantization axis for the levels $|F,m_F\rangle$.

The magnetic component of the microwave field couples the $|1,m_1\rangle$ to the $|2,m_2\rangle$ sublevels, with Rabi frequencies
\begin{equation} \label{eq:OmegaR}
\Omega_{1,m_1}^{2,m_2} (\mathbf{r}) = \frac{\langle 2,m_2| \hat{\boldsymbol{\mu}} \cdot \mathbf{B}_\mathrm{mw}(\mathbf{r}) |1,m_1\rangle }{\hbar},
\end{equation}
for the different transitions. We apply the rotating-wave approximation, which introduces negligible error for the parameters considered in this paper. In (\ref{eq:OmegaR}), $\hat{\boldsymbol{\mu}}=\mu_B g_J \mathbf{\hat J}$ is the operator of the electron magnetic moment ($g_J\simeq 2$); its matrix elements are evaluated in the basis $|F,m_F\rangle$ defined with respect to the local quantization axis along $\mathbf{B}_0(\mathbf{r})$. In a combined static magnetic and microwave trap, as considered here, both $\mathbf{B}_0(\mathbf{r})$ and $\mathbf{B}_\mathrm{mw}(\mathbf{r})$ vary with position. This leads to a position-dependent microwave coupling with in general all polarization components present. The detuning of the microwave from the resonance of the transition $|1,m_1\rangle \rightarrow |2,m_2\rangle$ is given by
\begin{equation}
\Delta_{1,m_1}^{2,m_2}(\mathbf{r}) = \Delta_0 - \frac{\mu_B}{2\hbar}(m_2 + m_1) |\mathbf{B}_0(\mathbf{r})|,
\end{equation}
where $\Delta_0 = \omega - \omega_0$ is the detuning from the transition $|F=1,m_F=0\rangle \rightarrow |F=2,m_F=0\rangle$, and the different Zeeman shifts (\ref{eq:VZ}) of the levels have been taken into account. In the following, we concentrate on the limit of large detuning $|\Delta_{1,m_1}^{2,m_2}|^2 \gg |\Omega_{1,m_1}^{2,m_2}|^2$, which allows for long coherence lifetimes of the qubit states in the microwave potential (see Sec.~\ref{ssec:CoherenceMWcoupl}). In this limit, the energy shifts due to the microwave coupling can be evaluated perturbatively for each transition. The overall magnetic microwave potential for the level $|F,m_F\rangle$ equals the sum of the energy shifts due to the individual transitions connecting to this level. Thus, the magnetic microwave potentials for the sublevels of $F=1$ are given by
\begin{equation}
V_\mathrm{mw}^{1,m_1}(\mathbf{r}) = \frac{\hbar}{4}\sum_{m_2}{ \frac{|\Omega_{1,m_1}^{2,m_2}(\mathbf{r})|^2} {\Delta_{1,m_1}^{2,m_2}(\mathbf{r})} },
\end{equation}
while the potentials for $F=2$ are given by
\begin{equation}
V_\mathrm{mw}^{2,m_2}(\mathbf{r}) = - \frac{\hbar}{4}\sum_{m_1}{ \frac{|\Omega_{1,m_1}^{2,m_2}(\mathbf{r})|^2} {\Delta_{1,m_1}^{2,m_2}(\mathbf{r})} }.
\end{equation}
As desired, the potentials for $F=1$ and $F=2$ have opposite sign, leading to a differential potential for the qubit states $|0\rangle$ and $|1\rangle$. In Fig.~\ref{fig:HFSmw}, the relevant transitions contributing to the potentials for $|0\rangle$ and $|1\rangle$ are shown for a general microwave field with all polarization components present. From Fig.~\ref{fig:HFSmw} it is evident that the microwave potentials for the two states will not have exactly the same position dependence (in addition to the difference in sign), since only two polarization components contribute to the potential for $|1\rangle$, while all three components contribute to the potential for $|0\rangle$. Furthermore, by reversing the sign of the detuning $\Delta_{1,m_1}^{2,m_2}$ for all levels, the overall sign of the microwave potentials and therefore the role of the states during the gate operation can be interchanged, i.e. either $|0\rangle$ or $|1\rangle$ can be chosen as the oscillating state. We choose $\Delta_{1,m_1}^{2,m_2}>0$, resulting in $|1\rangle$ as the oscillating state, since the potential for $|1\rangle$ is closer to harmonic in our actual trap design.

In addition to the magnetic microwave field, the electric field $\mathbf{E}_\mathrm{mw}(\mathbf{r})\cos(\omega t + \varphi)$ also leads to energy shifts. The electric field of the microwave polarizes the atoms, leading to a time-averaged quadratic Stark shift
\begin{equation}\label{eq:Vel}
V_\mathrm{el} (\mathbf{r}) = -\frac{\alpha}{4} |\mathbf{E}_\mathrm{mw}(\mathbf{r})|^2,
\end{equation}
where $\alpha$ is the scalar DC polarizability of the $^{87}$Rb ground state. In (\ref{eq:Vel}), we have averaged over the fast oscillation of the microwave at frequency $\omega$, which is much faster than the atomic motion. Eq.~(\ref{eq:Vel}) is actually the DC limit of optical dipole potentials \cite{Engler00}. Since all levels $|F,m_F\rangle$ belong to the electronic ground state, their polarizabilities are equal, leading to an electric microwave potential which is the same for all levels.

The total microwave potential for state $|0\rangle$ is identified with $u_0(\mathbf{r})$ in (\ref{eq:StateSelPot}),
\begin{eqnarray}
u_0(\mathbf{r}) &=& V_\mathrm{el} (\mathbf{r}) + V_\mathrm{mw}^{1,-1}(\mathbf{r}) \\ \nonumber
&=& - \frac{\alpha}{4} |\mathbf{E}_\mathrm{mw}(\mathbf{r})|^2
+ \frac{\hbar}{4}\sum_{m_2=-2}^{0}{ \frac{|\Omega_{1,-1}^{2,m_2}(\mathbf{r})|^2} {\Delta_{1,-1}^{2,m_2}(\mathbf{r})} },
\end{eqnarray}
while the microwave potential for state $|1\rangle$ is identified with $u_1(\mathbf{r})$ in (\ref{eq:StateSelPot}),
\begin{eqnarray}
u_1(\mathbf{r}) &=& V_\mathrm{el} (\mathbf{r}) + V_\mathrm{mw}^{2,+1}(\mathbf{r}) \\ \nonumber
&=& - \frac{\alpha}{4} |\mathbf{E}_\mathrm{mw}(\mathbf{r})|^2
- \frac{\hbar}{4}\sum_{m_1=0}^{+1}{ \frac{|\Omega_{1,m_1}^{2,+1}(\mathbf{r})|^2} {\Delta_{1,m_1}^{2,+1}(\mathbf{r})} }.
\end{eqnarray}


\section{Microwave chip design \label{sec:ChipDesign}}

\begin{figure}
\includegraphics[scale=0.55]{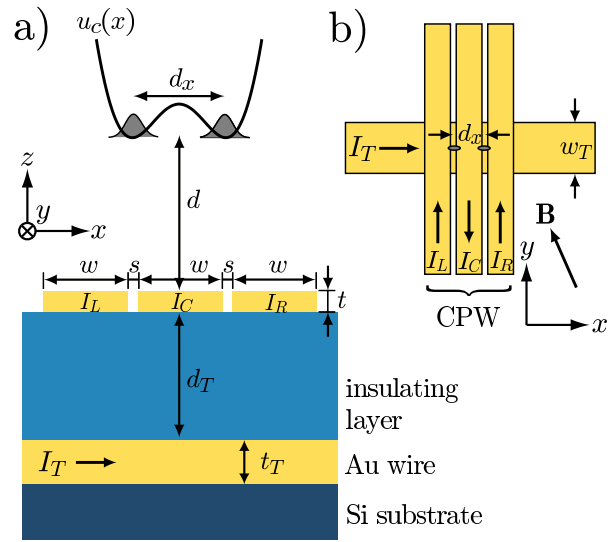}
\caption{\label{fig:WireLayout} (Color online) Chip layout for the microwave collisional phase gate. (a) Cut through substrate. (b) Top view of wire layout. Wire parameters: $w=0.8\,\mu$m, $s=0.1\,\mu$m, $t=0.2\,\mu$m, $d_T=4.2\,\mu$m, $w_T=1.5\,\mu$m, $t_T=1.0\,\mu$m. The lower wire (current $I_T$) is fabricated into a groove which was etched into the substrate, and covered by an insulating layer. The DC currents $I_T$, $I_C$, $I_L$, and $I_R$ and the orientation of the magnetic bias field $\mathbf{B}$ used to create the state-independent potential $u_c(\mathbf{r})$ are shown. The atomic wave functions in this potential are indicated. The three upper wires form a microwave coplanar waveguide (CPW), compare Fig.~\ref{fig:MWfields}.}
\end{figure}

To realize the state-selective potential for the atoms, we propose to use a microwave atom chip with a wire layout as shown in Fig.~\ref{fig:WireLayout}. The chip consists of two layers of gold metallization on a high resistivity silicon substrate, separated by a thin dielectric insulation layer. The wires carry stationary (DC) currents, which, when combined with appropriate stationary and homogeneous magnetic bias fields, create the state-independent potential $u_c(\mathbf{r})$. In addition to carrying DC currents, the three wires on the upper gold layer form a coplanar waveguide (CPW) \cite{Collin01} for a microwave at frequency $\omega$. The microwave fields guided by these conductors create the state-dependent potential $u_i(\mathbf{r})$. The combination of DC and microwave currents in the same wires is possible by the use of bias injection circuits \cite{Rizzi88}.

\subsection{Static magnetic trap}

For the phase gate with three oscillations ($N=3$, see Sec.~\ref{sec:Optimization}), an initial static trap with DC currents $I_T = 348.440$\,mA, $I_C = -0.813$\,mA, and $I_L = I_R = 1.204$\,mA is used (see Fig.~\ref{fig:WireLayout}). The components of the homogeneous bias field $\mathbf{B} = (B_x,B_y,B_z)$ are $B_x = -4.464$\,G, $B_y = 103.717$\,G, and $B_z = 0.000$\,G. For assumptions about the stability and accuracy of the currents and magnetic fields, see Sec.~\ref{ssec:TechLimit}.
In the calculation of the potential, the finite size of the wires is taken into account. For the parameters given above, $u_c(\mathbf{r})$ is a Ioffe-type double well potential \cite{Reichel02} along an axis $x^\prime$ in the $xy$-plane, which is tilted by a small angle $\theta=0.02$ with respect to $x$. The distance of the double well from the wire surface is $d=1.80\,\mu\mathrm{m}$. The magnetic field in the trap center is $|\mathbf{B_0}| = 3.230\,\textrm{G}$, which maximizes the coherence time in the absence of microwave coupling \cite{Harber02}. $\mathbf{B_0}$ is directed approximately along $-x$. The distance between the minima of the double well is $d_x = 1.32\,\mu\textrm{m}$. In the transverse dimensions, the trap provides a tight harmonic confinement with almost identical trap frequencies $\omega_y/2\pi = \omega_z/2\pi = \omega_\perp/2\pi = 77.46\,\textrm{kHz}$. For the simulation, we determine $\omega_\perp(x)$ as a function of the longitudinal coordinate $x$, although the relative variation along $x$ is only $\leq 1 \times 10^{-3}$. The axial trap frequency of the potential wells is $\omega_x/2\pi = 4.432\,\textrm{kHz}$. All traps considered in our gate simulation satisfy $\omega_x \ll \omega_\perp$.

\subsection{Microwave propagation on the chip \label{ssec:MWprop}}

To accurately determine the state-selective part of the potential, a simulation of the microwave electromagnetic fields in the proximity of the conductors transmitting the microwave signal is necessary. Since the transverse ($xz$-plane) size of the CPW is only a few micrometers, much smaller than the centimeter wavelength of the microwave, a quasi-static field analysis is sufficient to determine the microwave propagation characteristics \cite{Collin01}. At $\omega/2\pi = 6.8$\,GHz, the skin depth is
$\delta_s = \sqrt{2/\omega \mu_0 \sigma} = 0.9\,\mu\textrm{m}$,
where $\sigma=4.5\times 10^7\,\Omega^{-1}\mathrm{m}^{-1}$ is the conductivity of gold. The skin depth has to be compared to the thickness $t$ of the CPW conductors, see Fig.~\ref{fig:WireLayout}. Since $t < \delta_s$, the microwave electromagnetic field is not screened from the inside of the conductors, and microwave currents flow in the whole cross section of the gold wires, not only in a thin layer below the conductor surface. Taking this into account is equivalent to a proper treatment of conductor losses in the microwave simulation. The common assumption of perfect conductor boundary conditions on the wires is no longer justified.
The quasistatic characteristics of a micron-sized CPW with a proper treatment of conductor-loss effects are given in \cite{Heinrich93,Kunze99}. Compared to conductor losses, dielectric losses are negligible if a high-resistivity Si substrate is used, which is a standard substrate material for microwave circuits.

\begin{figure*}[]
  \includegraphics[scale=0.65]{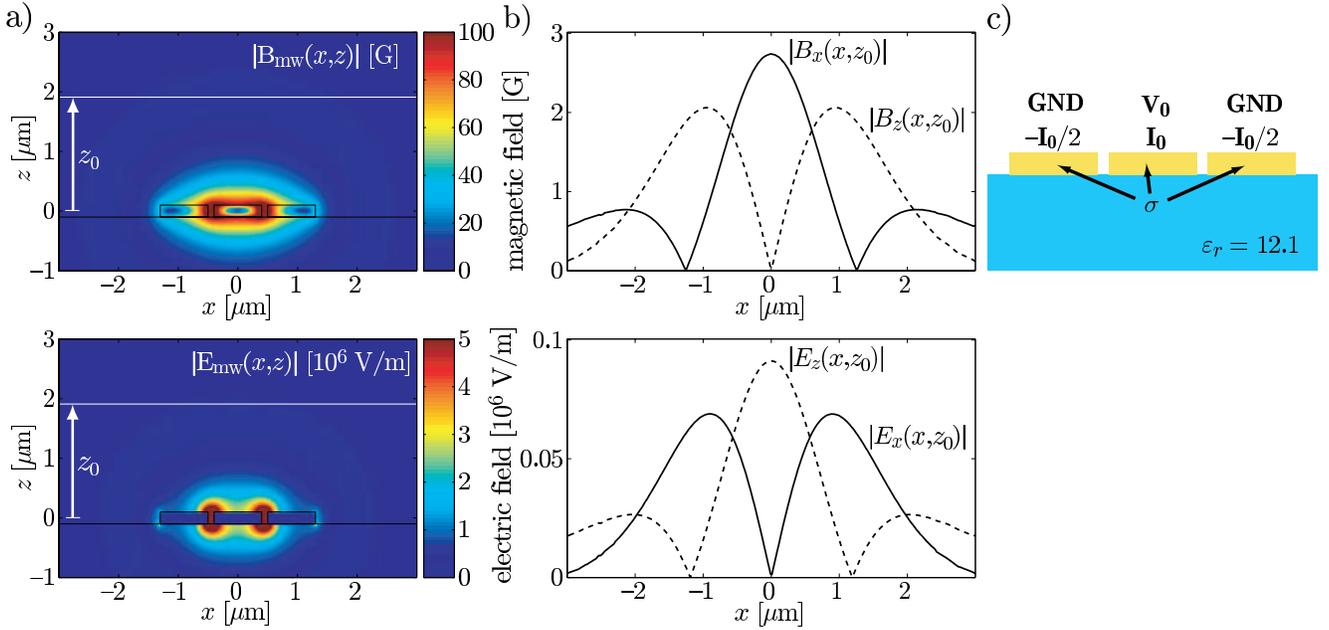}
\caption{\label{fig:MWfields} (Color online) Transverse ($xz$-plane) magnetic and electric fields of the microwave on the coplanar waveguide. The fields were obtained by a quasi-static simulation including conductor-loss effects. The longitudinal ($y$-direction) electric field, which is several orders of magnitude smaller than the transverse electric field, is not shown in the plots. (a) Magnitude of the fields for $\lambda=1$ shown in a cross section of the CPW. (b) Transverse microwave field components as a function of $x$ at a distance $d=1.80\,\mu\mathrm{m}$ from the wire surface, corresponding to the line $z=z_0=1.90\,\mu$m in (a), which indicates the $z$-position of the static trap minimum. (c) Microwave voltage and current amplitudes on the CPW used in the simulation. $\sigma$ is the conductivity of the gold wires, $\epsilon_r$ the dielectric constant of the Si substrate. For the size of the conductors, see Fig.~\ref{fig:WireLayout}.
}
\end{figure*}

We have performed a quasistatic analysis of the CPW in Fig.~\ref{fig:WireLayout}, neglecting the wire in the lower plane of metallization. The analysis consists of an electrostatic and a magneto-quasistatic simulation of the fields in a two-dimensional transverse ($xz$-plane) cross section of the waveguide. The results of this simulation are shown in Fig.~\ref{fig:MWfields}. From the fields, the complex characteristic impedance $\mathbf{Z_c}$ of the CPW can be determined. For our CPW,
$\mathbf{Z_c} = 130\,\Omega \cdot e^{-i\cdot 0.24\, \pi}.$
The microwave propagation constant $\beta_\mathrm{mw} = 2 \pi /\lambda_\textrm{mw}$ and the attenuation constant $\alpha_\mathrm{mw}$ are of comparable magnitude, $\beta_\mathrm{mw} \sim \alpha_\mathrm{mw} \sim 1\times 10^3\,\mathrm{m}^{-1}$ at $6.8$\,GHz. Damping and propagation effects are negligible on a micron scale.
The influence of the lower wire on the microwave propagation characteristics of the CPW is expected to be negligible, since $d_T \gg s$. To estimate the effect of the lower wire, we have included a homogeneous gold layer in the plane of the lower wire in our two-dimensional simulation. This leads to a small relative change in $|\mathbf{Z_c}|$ of $1 \times 10^{-2}$.

The microwave signal on the CPW is given by a complex voltage amplitude $\mathbf{V_0}$ on the center conductor, while the two outer conductors are set to $\mathbf{V}=0$ (ground), see Fig.~\ref{fig:MWfields}. Here we use phasor notation to describe the microwave signal and suppress the harmonic time dependence $e^{i\omega t}$ \cite{Collin01}. The amplitude of the microwave current in the center conductor is $\mathbf{I_0}$, while the two outer conductors carry a current $-\mathbf{I_0}/2$. Voltage and current are related by $\mathbf{V_0} = \mathbf{Z_c} \mathbf{I_0}$, i.e. there is a phase shift $\varphi = \arg(\mathbf{Z_c})$ between them, which gives rise to the phase shift $\varphi$ between electric and magnetic field of the microwave. The microwave potentials, however, are independent of $\varphi$ (see Sec.~\ref{ssec:MWpot}).

In the simulation of the gate dynamics and in Fig.~\ref{fig:MWfields}, the amplitudes of the microwave signal on the coplanar waveguide are
\begin{eqnarray}
|\mathbf{V_0}| &=& \sqrt{\lambda(t)} \cdot 1.9895\,\textrm{V} \\
|\mathbf{I_0}| = \frac{|\mathbf{V_0}|}{|\mathbf{Z_c}|} &=&  \sqrt{\lambda(t)} \cdot 15.343\,\textrm{mA}
\end{eqnarray}
Here, $\lambda(t)$ is the modulating function of the microwave potential during the gate operation, which will be optimized by optimal control techniques as described in Sec.~\ref{sec:Optimization}. The detuning of the microwave from the transition $|F=1,m_F=0\rangle \rightarrow |F=2,m_F=0\rangle$ is
$\Delta_0 = 2\pi \cdot 29.4\,\textrm{MHz}$.
For these parameters, the ratio between the electric and the magnetic microwave potential is $V_\textrm{el} \sim 0.2\,V_\textrm{mw}$, i.e. the state-selective magnetic part of the microwave potential is dominating. This is important since $V_\textrm{el}$ reduces the barrier of the double well also for state $|0\rangle$. It is possible to avoid this unwanted effect of $V_\textrm{el}$ by a small adjustment of the magnetostatic potential during the gate operation. To compensate $V_\textrm{el}$, we modulate
\begin{eqnarray}
B_x(t) &=& (-4.464 + \lambda_0(t) \cdot 0.036)\,\textrm{G} \\
I_C(t) &=& (-0.813 - \lambda_0(t) \cdot 0.039)\,\textrm{mA},
\end{eqnarray}
where $\lambda_0(t) \simeq \lambda(t)$. The function $\lambda_0(t)$ consists of simple linear ramps, as explained below (Sec.~\ref{ssec:OptimLambda}).

If the microwave is turned on to full power ($\lambda=1$ and $\lambda_0=1$), the barrier of the static double well trap is removed for state $|1\rangle$, leaving a single potential well of longitudinal trap frequency $\omega_{|1\rangle}/2\pi = 5.448$\,kHz, as shown in Fig.~\ref{fig:StateSelPot}. For state $|0\rangle$, the two potential wells are shifted apart and the trap frequency in each of the wells changes to $\omega_{|0\rangle}/2\pi = 4.775$\,kHz. The transverse confinement of the static trap is unchanged, both $d$ and $\omega_\perp$ change by less than $10^{-3}$. Although we include the full position dependence of $u_i(\mathrm{r})$ in the simulation, it would be sufficient to consider the state-selective potential $u_i(x,y_0,z_0)$ at the transverse position of the static trap minimum $(y_0,z_0)$.

\subsection{Chip fabrication}

The fabrication of the chip relies on electron beam lithography, which provides sub-micron resolution, in combination with lift-off metallization techniques and anisotropic etching of the Si substrate. The lower gold wire is fabricated as follows: prior to the metal deposition, a groove is etched into a high-resistivity Si substrate by anisotropic reactive ion etching, using lithographically patterned photoresist as etch mask. The gold wire is subsequently fabricated into this groove using the same photoresist for lift-off. The remaining groove is filled and the wire is insulated by depositing a dielectric layer, such as SiO$_2$ or polyimide. Since the lower wire is contained in the groove, this insulating layer can be very thin and still provide good planarization of the surface. In this way, good thermal contact of the wires to the substrate is guaranteed. On top of the insulating layer, the CPW is fabricated using standard electron-beam lithography and lift-off.


\section{Simulation of the gate operation}

The gate operation is simulated numerically by solving a time-dependent Schr{\"o}dinger equation of the two-particle dynamics along the direction of the double well $x^\prime$ (in the following, we write $x$ instead of $x^\prime$ to simplify notation). We assume that the atoms remain in the transverse ground state of the trap throughout the gate operation; this assumption is justified in Sec.~\ref{ssec:OptimalOmegaPerp} and Sec.~\ref{ssec:quasi1D}. The initial state of the two atomic qubits in an arbitrary superposition of internal states $|ij\rangle$ is
\begin{equation}
|\Psi(t=0)\rangle = \left( \sum_{i,j \in \{0,1\}} \alpha_{ij} |ij\rangle \right) \otimes |\psi(x_1,x_2,t=0)\rangle,
\end{equation}
where $|\psi(x_1,x_2,t=0)\rangle$ describes the initial motional state of the two atoms, which is independent of the internal state, with one atom in each well of the potential $u_c(x)$ (see Fig.~\ref{fig:StateSelPot}a). During the gate operation, the internal and motional states of the atoms are entangled due to the dynamics in the state-selective potential:
\begin{equation}
|\Psi(t)\rangle = \sum_{i,j \in \{0,1\}} \alpha_{ij} |ij\rangle \otimes |\psi_{ij}(x_1,x_2,t)\rangle.
\end{equation}
The dynamics of $|\psi_{ij}(x_1,x_2,t)\rangle$ (see Fig.~\ref{fig:StateSelPot}b) is governed by the time-dependent Hamiltonian
\begin{equation}
\mathcal{H}_{ij}(t) = {T}_{x_1} + {T}_{x_2} + U_i(x_1,t) + U_j(x_2,t) + V^{ij}_\mathrm{int}(|x_2 - x_1|,t),
\end{equation}
where $T_{x_i}$ denotes the kinetic energy operators and
\begin{equation}\label{eq:Vint}
V^{ij}_\mathrm{int}(|x_2 - x_1|,t) = \frac{2\hbar\omega_\perp(x_1,t) a^{ij}_s}{1-1.46a^{ij}_s/a_{\perp}(x_1,t)} \delta(|x_2 - x_1|)
\end{equation}
is the effective one-dimensional interaction potential between the atoms \cite{Olshanii98},  $a^{ij}_s \simeq 5.4$~nm is the s-wave scattering length for collisions of $^{87}$Rb atoms in state $|ij\rangle$, and $a_{\perp}(x_1,t) = \sqrt{2 \hbar/m \omega_\perp(x_1,t)}$ is the size of the ground state in the transverse direction. The transverse trap frequency $\omega_\perp(x_1,t)$ can be used as a control parameter to optimize the dynamics and is therefore time-dependent. The numerical simulation of the dynamics is performed
using a Fast Fourier Transform method.

The entanglement between the motional and internal states of the atoms is crucial for establishing the collisional phase shift in state $|11\rangle$. However, at the end of the gate operation, the motional and internal states have to factorize again, so that the truth table (\ref{eq:TruthTable}) is implemented. Any residual entanglement between the motional and internal states leads to a reduction of the gate fidelity. To avoid this, the wave function of the atoms has to show a revival at $t=\tau_g$ and regain its initial form, apart from the phase factor in state $|11\rangle$. To quantify this, we define the overlaps
\begin{equation}
O_{ij}(t) = \langle \psi_{ij}(x_1,x_2,t)|\psi(x_1,x_2,0)\rangle,
\end{equation}
from which the overlap fidelities
\begin{equation}
F_{ij}(t) = |O_{ij}(t)|^2
\end{equation}
are calculated.
The evolution of the gate phase is given by \cite{Calarco00,Calarco01}
\begin{equation}
\phi_g(t) = \phi_{11}(t) + \phi_{00}(t) - \phi_{01}(t) - \phi_{10}(t),
\end{equation}
where we have defined the collisional phases
\begin{equation}
\phi_{ij}(t) = \arg\left[\langle \psi_{ij}(x_1,x_2,t)|\psi^0_{ij}(x_1,x_2,t)\rangle\right].
\end{equation}
Here, $|\psi^0_{ij}(x_1,x_2,t)\rangle$ is the motional state evolved from $|\psi(x_1,x_2,0)\rangle$ in the absence of collisions \cite{Calarco00}. $\phi_{ij}(t)$ is well defined even for times $t$ where $O_{ij}(t)\simeq 0$.
Optimal gate performance corresponds to $F_{ij}(\tau_g) = 1, \forall_{i,j}$, i.e. a complete revival of the wave function, and a gate phase $\phi_g(\tau_g) = \pi$. In a realistic non-harmonic potential, however, the revival of the atomic wave function will be incomplete, and the collisions between the atoms will lead to an additional distortion of the wave function. Both effects will lead to a reduction of the gate fidelity, and have to be avoided as much as possible by an optimization of the gate dynamics.

\section{Optimization of the gate\label{sec:Optimization}}

We apply optimal control techniques \cite{Sklarz02,Calarco04} to optimize the gate performance. This is performed in two steps. First, the dynamics without atom-atom interactions is optimized using $\lambda(t)$ as control parameter. In this way, we optimize the revival of the wave function in the non-harmonic potential.
In a second step, the dynamics in the presence of interactions is optimized further using $\omega_\perp(t)$ as control parameter. This reduces wave packet distortion due to collisions. The two steps are discussed in the following subsections for a gate with $N=3$ oscillations.

\subsection{Optimal control of $\lambda(t)$ \label{ssec:OptimLambda}}

\begin{figure*}
\includegraphics[scale=0.6]{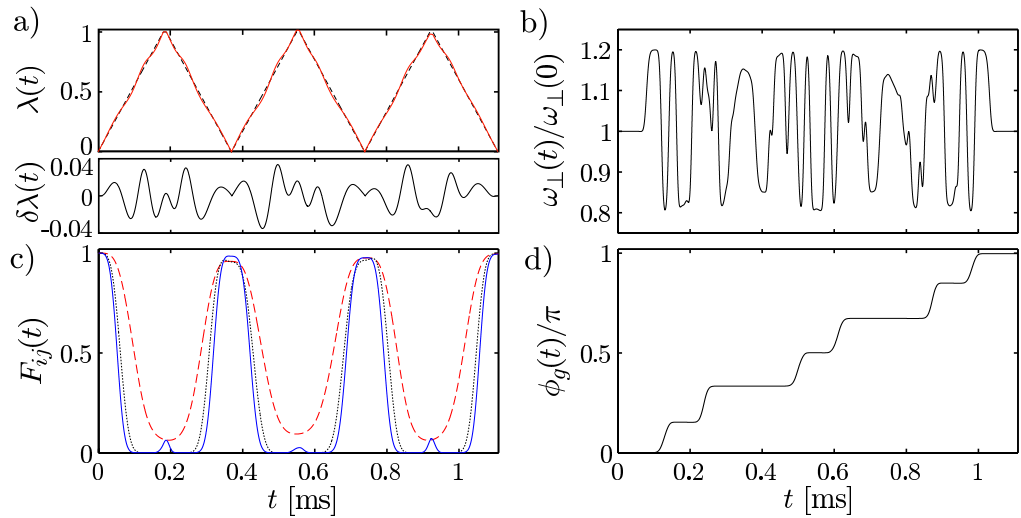}
\caption{\label{fig:Results} (Color online) Dynamics during the gate operation, shown for $N=3$ oscillations, the gate time is $\tau_g=1.110$\,ms. Both $\lambda(t)$ and $\omega_\perp(t)$ are used as control parameters. (a) Optimal control of the microwave power. Upper plot: Initial trial function $\lambda_0(t)$ (dashed) and optimized control parameter $\lambda(t)$ (solid line). Each triangular ramp of $\lambda(t)$ corresponds to a full oscillation of state $|1\rangle$. Lower plot: the difference $\delta\lambda(t) = \lambda(t)-\lambda_0(t)$ shows small modulations. (b) Optimal control of the effective one-dimensional interaction strength via modulation of the transverse trap frequency $\omega_\perp(t)$. (c) Evolution of the overlap fidelities during the gate operation: $F_{00}(t)$ (dashed), $F_{01}(t)=F_{10}(t)$ (dotted), $F_{11}(t)$ (solid line). (d) Evolution of the gate phase $\phi_g(t)$. The phase shift steps are due to the six collisions in state $|11\rangle$.}
\end{figure*}

In the first stage of optimization, $\lambda(t)$ is used as control parameter. Experimentally, this corresponds to controlling the microwave power. We start with a trial function $\lambda_0(t)$ which consists of linear ramps, as shown in Fig.~\ref{fig:Results}a. This choice represents a compromise between a sudden removal of the barrier for state $|1\rangle$ and an adiabatic shift of the potential wells for state $|0\rangle$. We optimize $\tau_g$ and the slope of the linear ramps of $\lambda_0(t)$ by hand to achieve an initial gate fidelity $>0.95$. Starting with $\lambda(t) = \lambda_0(t)$, the function $\lambda(t)$ is then adjusted by an optimal control algorithm, which neglects atom-atom interactions in order to provide faster convergence. This corresponds to setting $a_s=0$ in (\ref{eq:Vint}). In the absence of interactions, it is sufficient to optimize the single-particle overlaps $O_i = |\langle \psi_i(x,\tau_g)|\psi(x,0)\rangle|$, which are optimized simultaneously for $i\in\{0,1\}$ by the algorithm using the Krotov method \cite{Krotov96}. After the optimization has converged, the resulting optimized $\lambda(t)$ is used in a simulation of the two-particle dynamics with interactions ($a_s\neq 0$), in order to determine the two-particle overlap fidelities $F_{ij}(\tau_g)$. The strength of the interaction (\ref{eq:Vint}) was adjusted by setting the transverse trap frequency $\omega_\perp$ to a value constant in time such that the gate phase is $\phi_g \simeq \pi$. The result of this first stage of optimal control is the microwave power modulation function $\lambda(t)$ shown in Fig.~\ref{fig:Results}a, which deviates from $\lambda_0(t)$ by small modulations, and the fidelities shown in the left part of Table~\ref{tab:Fidelities}.

The trial function $\lambda_0(t)$ is also used to modulate the magnetostatic potential in order to compensate the effect of the microwave electric field, as explained in Sec.~\ref{ssec:MWprop}. We choose $\lambda_0(t)$ instead of $\lambda(t)$ for this modulation in order to keep the number of experimental parameters which are subject to optimal control as small as possible.

Our simulation shows that collisional interactions between the atoms are negligible in basis states other than $|11\rangle$, with $\phi_{00}(\tau_g)=0$ and $\phi_{01}(\tau_g)=\phi_{10}(\tau_g) \sim 10^{-3}\, \phi_{11}(\tau_g)$. In Fig.~\ref{fig:Results}d, the phase evolution during the gate operation is shown. The main effect of the interaction is to provide the phase shift in state $|11\rangle$, corresponding to the steps in $\phi_g(t)$ each time a collision in this state takes place. An undesired effect of the collision is to decrease the fidelity $F_{11}(\tau_g)$ compared to the non-interacting case due to wave packet distortion during the collision.
We have performed the optimization of the gate for different numbers of oscillations $N$. For each value of $N$, a different value of $\omega_\perp$ was chosen to adjust the gate phase to $\phi_g\simeq\pi$. For smaller $N$, the gate phase has to be acquired in a smaller number of oscillations, therefore the interactions need to be stronger and the phase shift per collision is larger. Correspondingly, the collisional distortion of the wave function is also larger, and $F_{11}$ is smaller. This general tendency can be seen in the left part of Table~\ref{tab:Fidelities} by comparing the values of $F_{11}$ for different $N$. The best gate performance was achieved for $N=5$, with $F_{11}=0.991$ and a gate time of $\tau_g = 1.838$\,ms.

\begin{table*}
\begin{tabular}{|c c|*{6}{c}|*{4}{c}|}

\multicolumn{2}{c}{} & \multicolumn{6}{c}{Optimal control of $\lambda(t)$} &  \multicolumn{4}{c}{Optimal control of $\lambda(t)$ and $\omega_\perp(t)$} \\
\hline
$N$ & $\tau_g$ [ms] & $O_0$ & $O_1$ & $F_{00}$ & $F_{01}$ & $F_{11}$ & $\phi_g/\pi$ & $F_{00}$ & $F_{01}$ & $F_{11}$ & $\phi_g/\pi$ \\
\hline
\hline
2 & 0.696 & 0.996 & 0.996 & 0.993 & 0.993 & 0.960 & 0.996 & 0.993 & 0.993 & 0.987 & 0.998  \\
3 & 1.110 & 0.998 & 0.997 & 0.996 & 0.996 & 0.986 & 0.999 & 0.996 & 0.996 & 0.995 & 0.997  \\
4 & 1.389 & 0.996 & 0.995 & 0.991 & 0.991 & 0.991 & 0.991 &  &  &  &   \\
5 & 1.838 & 0.999 & 0.997 & 0.997 & 0.997 & 0.991 & 0.995 &  &  &  &   \\
6 & 2.219 & 0.996 & 0.998 & 0.992 & 0.993 & 0.985 & 0.993 &  &  &  &   \\
\hline
\end{tabular}
\caption{\label{tab:Fidelities} Optimized gate performance for different numbers of oscillations $N$ and correspondingly different gate times $\tau_g$. All two-particle fidelities $F_{ij}$, single-particle overlaps $O_i$, and the gate phase $\phi_g$ are evaluated at $t=\tau_g$. Either only $\lambda(t)$ (left part of the table) or both $\lambda(t)$ and $\omega_\perp(t)$ (right part of the table) were chosen as control parameters. The optimization of $\omega_\perp(t)$ was performed only for $N\leq 3$, which is particularly interesting due to the short gate times $\tau_g \leq 10^{-3}\, \tau_t$.
}
\end{table*}


\subsection{Optimal control of $\omega_\perp(t)$ \label{ssec:OptimalOmegaPerp}}

After optimization of $\lambda(t)$, the fidelity of the gate is limited by collisional distortion of the wave function in state $|11\rangle$. To overcome this limitation, we have implemented a second stage of optimization, in which the interaction potential (\ref{eq:Vint}) is controlled during the gate operation with the transverse trap frequency $\omega_\perp(t)$ as control parameter. Experimentally, this corresponds to controlling the bias field component $B_y(t)$ and the DC current $I_T(t)$ proportional to the desired modulation of $\omega_\perp(t)$.
Since the collisions in internal states other than $|11\rangle$ are negligible, this optimization affects only state $|11\rangle$. We optimize the overlap fidelity $F_{11}(\tau_g)$ and the phase $\phi_{11}(\tau_g)$ simultaneously, using again the Krotov method. For the microwave power we use the optimized $\lambda(t)$ determined in the previous section.

Care has to be taken that the modulation of $\omega_\perp(t)$ does not create excitations of the transverse state of the atoms. Since it is not possible to put constraints on the time derivative of the control parameter in the Krotov method of optimization, we have chosen the following strategy to avoid transverse excitations. We parameterize the transverse trap frequency
\begin{equation}
\omega_\perp(t) = \omega_\perp(0) [A \tanh \alpha(t) + 1],
\end{equation}
with $\alpha(t)$ being the dimensionless optimal control parameter, and $\omega_\perp(0)$ is determined for a given $N$ as in the previous section. In this way, the optimization is constrained to a maximum modulation amplitude set by $A$. After the optimization on $\alpha(t)$ has converged, we remove high-frequency components from the resulting modulation of $\omega_\perp(t)$ by filtering with a cutoff frequency $\omega_c$. We choose $\omega_c < 2\,\omega_\perp(0)$ in order to avoid parametric excitation of the transverse degrees of freedom.
By performing the gate simulation again with the filtered modulation $\omega_\perp(t)$, we determine the overlap fidelities and the gate phase. We have seen that the filtering does not significantly decrease the fidelity if $\omega_c \gg \omega_x$. This shows that the high frequency components with $\omega > \omega_c$ were artifacts of the optimization algorithm without physical significance.
The modulation of $\omega_\perp(t)$ after filtering is shown in Fig.~\ref{fig:Results}b.
For this modulation, $A=0.2$ and $\omega_c = 0.8\, \omega_\perp(0)$. To quantify the transverse excitation probability $p^m_\perp(t)$ during the gate operation, we simulate the transverse dynamics using a harmonic oscillator model, where the frequency is modulated according to the filtered $\omega_\perp(t)$. This simulation yields $p^m_\perp(t) < 7 \times 10^{-4}$, therefore transverse excitations due to the modulation of $\omega_\perp(t)$ do not limit the gate performance.

Fig.~\ref{fig:Results} is the main result of our simulation. It shows the gate dynamics for $N=3$ after optimization of both $\lambda(t)$ and $\omega_\perp(t)$. Corresponding numbers for the overlap fidelities $F_{ij}$ and the gate phase $\phi_g$ are shown in the right part of Table~\ref{tab:Fidelities}. The improvement compared to optimal control of $\lambda(t)$ alone (left part of Table~\ref{tab:Fidelities}) is an increase of $F_{11}$ from $F_{11}=0.986$ to $F_{11}=0.995$, which is now comparable to the overlap fidelities in the other states. The gate time is $\tau_g = 1.110$\,ms.


\subsection{Gate fidelity}

In order to provide an estimate of the overall gate performance, we use the definition given in \cite{Calarco00} for the gate fidelity:
\begin{eqnarray}
F={\rm min}_{\chi}\left\{{\rm Tr}_{{\rm ext}}\left[\langle\tilde{\chi}|\mathcal{U}\,\mathcal{S}\left(
|\chi\rangle\langle\chi|\otimes\rho_0\right)\mathcal{S}^{\dagger}\mathcal{U}^{\dagger}|\tilde{\chi}\rangle\right]\right\}.
\end{eqnarray}
Here, $|\chi\rangle$ is an arbitrary internal state of both atoms, and $|\tilde{\chi}\rangle$ is the state resulting from $|\chi\rangle$ using the actual transformation. $\mathcal{U}$ and $\mathcal{S}$ are the operators for time evolution and symmetrization under particle interchange, respectively. The density matrix $\rho_0$ is the initial two-particle motional state.
According to this definition, the gate fidelity is
\begin{equation}\label{eq:finalfidelity}
F=0.997 \quad \textrm{for} \quad N=3,\, \tau_g = 1.110\,\textrm{ms}
\end{equation}
after optimization of both $\lambda(t)$ and $\omega_\perp(t)$. Even after the optimization, wave packet distortion still contributes the largest error reducing the fidelity. The error sources discussed in the next section lead to an additional error of the order of $1 \times 10^{-3}$, mainly due to the finite trap lifetime, which is limited by surface effects. If we include this error in the calculation of the fidelity, we get $F=0.996$. By comparison, with trap and coherence lifetimes $\tau_t \sim \tau_c \sim 10^3\,\tau_g$, the maximum achievable fidelity of the gate with $N=3$ would be $F=0.999$, if wave packet distortion due to interaction dynamics could be reduced to negligible values.


\section{Error sources}

In this section, we discuss several effects which could possibly limit the fidelity of the phase gate proposed here, and justify the assumptions made in the description of the gate dynamics.

\subsection{Finite temperature}

\begin{figure}
\includegraphics[scale=0.4]{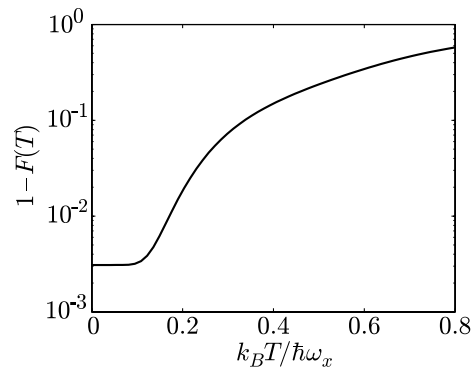}
\caption{\label{fig:FT} Gate infidelity $1-F(T)$ as a function of temperature for the gate with $N=3$.}
\end{figure}

The result for the fidelity (\ref{eq:finalfidelity}) is obtained for zero temperature $(T=0)$ of the atoms. This corresponds to an initial motional state with one atom in the ground state $|n_1 = 0\rangle$ of the left potential well and the other atom in the ground state $|n_2 = 0\rangle$ of the right potential well of $u_c(\mathbf{r})$. To study the effect of finite temperature $T$ on the gate performance, we start with an initial density operator for the motional degrees of freedom
\begin{eqnarray}
\rho_0=\sum_{n_1,n_2 = 0}^\infty P_{n_1,n_2}(T)|n_1\rangle\langle n_1|\otimes|n_2\rangle\langle n_2|,
\end{eqnarray}
where $P_{n_1,n_2}(T)$ is the probability for the occupation of the motional state $|n_1\rangle$ for atom $1$ and $|n_2\rangle$ for atom $2$ in the initial double well trap. The probabilities $P_{n_1,n_2}(T)$ are calculated assuming a thermal distribution corresponding to a temperature $T$ in the canonical ensemble. We have evaluated the fidelity $F=F(T)$ of our gate with $N=3$ as a function of $T$, see Fig.~\ref{fig:FT}. The results show that high fidelity gate operations are only possible if the two atoms can be prepared in the motional ground state of the trap, with very low occupation probability of excited states. We assume that $k_B T / \hbar \omega_x \leq 0.1$ can be reached, corresponding to a temperature $T \leq 20$\,nK in the initial double well trap. In this case, the fidelity is not reduced significantly due to the finite temperature.

\subsection{Condition for quasi-1D dynamics \label{ssec:quasi1D}}

To ensure one-dimensional dynamics of the atoms in the experiment and to justify the one-dimensional simulation of the gate presented above, transverse excitations due to the collisions in state $|11\rangle$ have to be avoided. This mechanism could limit the fidelity even if transverse excitations due to the modulation of the transverse trap frequency $\omega_\perp(t)$ are suppressed (cf. Sec.~\ref{ssec:OptimalOmegaPerp}).

\begin{figure}
\includegraphics[scale=0.4]{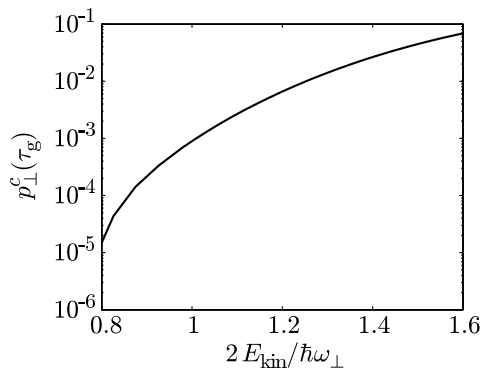}
\caption{\label{fig:TransEx} Transverse excitation probability $p^c_\perp(\tau_g)$ due to collisions of the atoms for $N=3$. $p^c_\perp(\tau_g)$ is shown as a function of the kinetic energy $E_\mathrm{kin}$ of one atom in state $|1\rangle$ at the time of the collision.}
\end{figure}

We calculate the probability $p^c_\perp(t)$ of transverse excitations due to collisions using a simple model. We consider two interacting atoms in a three-dimensional anisotropic harmonic oscillator potential, with $\omega_x \ll \omega_y = \omega_z \equiv \omega_\perp$. Exact solutions for this model are known \cite{Idziaszek05}. The atoms are initially in the transverse ground state of the trap, but separated in two harmonic potential wells with separation $d_x$ along the $x$ direction, similar to the initial state of our phase gate. At time $t=0$, the initial double well is switched off and the atoms evolve in the single anisotropic harmonic oscillator potential. The dynamics is obtained by expanding the initial three-dimensional two-atom state on the interacting basis given in \cite{Idziaszek05}. For each evolution time $t$, we compute the reduced density matrix of the transverse motion by tracing out the axial degrees of freedom. In this way we compute the probability $p^c_\perp(t)$ as a function of time. We have performed this calculation for different $d_x$, corresponding to different kinetic energies $E_\textrm{kin}$ of one atom at the time of the collision. In Fig.~\ref{fig:TransEx} we show $p^c_\perp(\tau_g)$ for $N=3$ as a function of $2\,E_\textrm{kin}/\hbar\omega_\perp$.
We find that transverse excitations are energetically suppressed if
\begin{equation}
2\,E_\textrm{kin} < \hbar \omega_\perp,
\end{equation}
For all trapping geometries used in our simulation of the phase gate,
$2\,E_\textrm{kin}/\hbar\omega_\perp \leq 0.7$,
and the transverse excitation probability is suppressed to negligible values.

\subsection{Microwave coupling and qubit dephasing\label{ssec:CoherenceMWcoupl}}

In contrast to optical dipole transitions, spontaneous emission is negligible for microwave transitions between ground state sublevels. Therefore microwaves can also be used to create adiabatic potentials with the microwave frequency tuned to resonance with a particular hyperfine transition, similar to the adiabatic radio-frequency potentials investigated in \cite{Zobay01,Colombe04,Schumm05}. However, resonant coupling results in strong mixing of the hyperfine levels connected by the microwave. This is undesirable for our qubit state pair, since it would destroy the excellent coherence properties of the qubit by admixtures of other states with different magnetic moments. For this reason, we concentrate on the limit of large microwave detuning in this paper. The large detuning also ensures that the modulation of the microwave power during the gate operation is adiabatic with respect to the internal state dynamics of the atoms.

For the trap parameters considered here,
\begin{equation}\label{eq:omdelta}
\max \frac{\Omega_R^2}{\Delta^2} \leq 10^{-2},
\end{equation}
where we have set $\Omega_R \equiv \Omega_{1,m_1}^{2,m_2}$ and $\Delta \equiv \Delta_{1,m_1}^{2,m_2}$ to shorten notation in this section.
In the absence of the microwave coupling, the differential magnetic moment $\delta \mu = \partial E_{|1\rangle} / \partial B - \partial E_{|0\rangle} / \partial B$ of the qubit states $|1\rangle$ and $|0\rangle$ can be calculated from the Breit-Rabi formula \cite{Harber02}. For magnetic fields $B_0 < 6$\,G, we get $| \delta\mu | < 2 \times 10^{-3}\,\mu_B$, with $\delta\mu = 0$, i.e. a vanishing first-order differential Zeeman shift, for $B_0 = 3.229$\,G \cite{Harber02}, as in the center of our static trap.
In the presence of the microwave coupling, $\delta\mu$ changes due to the admixture of other magnetic sublevels, whose magnetic moments differ by multiples of  $\mu_B/2$ from the magnetic moment of the qubit states.
For $\Omega_R^2 \ll \Delta^2$, the order of magnitude of this change can be estimated to
\begin{equation}\label{eq:deltamu}
\delta\mu \simeq \frac{\Omega_R^2}{4\Delta^2}\mu_B \leq 2.5 \times 10^{-3} \mu_B,
\end{equation}
where we have used (\ref{eq:omdelta}).

The differential magnetic moment determines the coherence time $\tau_c$ of the qubit in the presence of longitudinal magnetic field fluctuations (pure 'dephasing' of the qubit). The frequency spectrum of the fluctuations plays an important role in this context: Low frequency fluctuations of the magnetic field are most harmful to the coherence time \cite{Stern90}, since they do not average out on the time scale $\tau_g$ of a single gate operation.
In contrast to magnetic near-field noise arising from the chip surface (see Sec.~\ref{ssec:SurfEffects}), which has a flat spectrum in the relevant frequency range \cite{Henkel03}, technical magnetic field noise typically increases towards low frequencies and dominates in the frequency range $0 < \omega < 10/\tau_g$ considered here.

The atom chip is surrounded by magnetic shielding, so that the residual fluctuations of the magnetic field are limited by the stability of the current sources used for the wires and bias magnetic field coils. We assume that fluctuations in the frequency range $0 < \omega < 10/\tau_g$ can be reduced to an r.m.s. amplitude of $\delta B \leq 0.01$\,mG, consistent with the stability of the current sources in our experiments.
Under this assumption, we expect a coherence time of the order of
\begin{equation}
\tau_c \sim \frac{\hbar}{\delta\mu \, \delta B} \sim 5\,\textrm{s},
\end{equation}
where we have used (\ref{eq:deltamu}). Comparing this value of $\tau_c$ with the trap and coherence lifetimes due to surface effects calculated in Sec.~\ref{ssec:SurfEffects}, we find that pure dephasing of the qubit does not limit the fidelity of the gate.

\subsection{Trap loss and decoherence due to the chip surface \label{ssec:SurfEffects}}
A fundamental source of trap loss, heating, spin- and motional decoherence on atom chips is magnetic near-field noise originating from thermal currents in the chip wires \cite{Henkel03}. For our state pair, pure spin dephasing arising from longitudinal magnetic near-field noise is negligible. The dominating surface effect is trap loss due to spin-flips induced by transverse magnetic field noise.
For the chip layout shown in Fig.~\ref{fig:WireLayout} and a distance $d=1.80\,\mu$m from the chip surface, we estimate an average spin-flip rate of $\Gamma_s = 0.9\,\textrm{s}^{-1}$, taking the finite thickness and width of the wires \cite{Zhang05} and the different matrix elements for spin flips of state $|0\rangle$ and $|1\rangle$ into account. This corresponds to a trap lifetime of
\begin{equation}
\tau_t = \Gamma_s^{-1} = 1.1\,\mathrm{s}.
\end{equation}
Compared with surface effects, loss rates due to collisions with background gas atoms are negligible. The rates for heating and motional decoherence due to magnetic near-field noise are comparable to $\Gamma_s$ \cite{Henkel03}, so that we expect an overall coherence lifetime $\tau_c \sim \tau_t$.
For $N=3$, the gate operation time is $\tau_g = 1.110$\,ms. The mentioned surface effects introduce an error of $1-\exp(-\Gamma_s \tau_g) = 1\times 10^{-3}$ in the gate operation. This error is smaller than the error due to wave packet distortion, and therefore does not significantly decrease the gate fidelity (\ref{eq:finalfidelity}). The error could be further reduced by reducing the wire thickness or using a trap at a slightly higher atom-surface distance.

\subsection{Two-photon transitions}

Another potential source of infidelities are two-photon transitions induced by the microwave. These can arise if more than one polarization component of the microwave is present, as it is in general the case in microwave near-fields. For $\Omega_{Ri}^2 \ll \Delta^2$, the two-photon coupling is characterized by an effective Rabi frequency $\Omega_\textrm{2ph}=\Omega_{R1}\Omega_{R2}/(2\Delta)$, where $\Omega_{Ri}$, $i\in\{1,2\}$, are the single-photon Rabi frequencies of the transitions involved \cite{Gentile89}. The detuning from two-photon resonance $\Delta_\textrm{2ph} = \mu_B B_0 / (2 \hbar)$ is given by the Zeeman splitting due to the static magnetic field $B_0$ in the trap center.
For our trap parameters
\begin{equation}
\max \frac{\Omega_\textrm{2ph}^2}{\Delta_\textrm{2ph}^2} \leq 2 \times 10^{-3},
\end{equation}
and two-photon transitions are suppressed by the large two-photon detuning.

\subsection{Technical limitations \label{ssec:TechLimit}}

Since the conductor dimensions on our chip are smaller than the skin depth, the microwave current density in the wires is approximately constant over the whole wire cross section. Therefore the resistance of the wires is given by the DC resistance and the maximum tolerable microwave current density is comparable to the maximum DC current density in wires of this size. In our simulation, we assume total (microwave + DC) current densities $j_\textrm{tot} \leq 1\times 10^{11} \, \textrm{Am}^{-2}$ in the wires forming the coplanar strip line. The DC current density in the lower wire is $j_\textrm{tot}=2\times 10^{11} \, \textrm{Am}^{-2}$.
Comparable current densities have been realized experimentally in \cite{Groth04} on Si substrates with a 20\,nm SiO$_2$ insulating layer.
Compared to ohmic loss in the chip conductors, dielectric loss in the substrate can be neglected for the frequency and structure size considered here \cite{Collin01}, as it is confirmed by our microwave simulation.

We have checked that the potentials used in our simulation are robust against current and magnetic field fluctuations. We assume a relative stability on the level of $10^{-5}$ for the static currents and fields. With available current sources and magnetic shielding, such a stability can be reached in experiments. The accuracy of the currents (magnetic fields) specified in Sec.~\ref{sec:ChipDesign} is assumed to be better than $1\,\mu$A ($1$\,mG).

In order to study the robustness of the gate against noise on the time-dependent control parameters, we have simulated the gate with white noise on $\lambda(t)$ and $\omega_\perp(t)$. The gate fidelity is reduced by $\sim 10^{-4}$ for relative r.m.s. noise amplitudes $n_a < 10^{-3}$ on the control parameters. For these values of $n_a$, the small modulations of the experimental parameters used for optimal control are well above the noise level, see Fig.~\ref{fig:Results}.

\subsection{Scattering length}

For an accurate theoretical description in one dimension, the 1D scattering length $a_{1D}(k)$ should be approximately independent of energy \cite{Moore03,Idziaszek06}
\begin{equation}
\frac{|a_{1D}(k) - a_{1D}(k=0)|}{a_{1D}(k)} \ll 1.
\end{equation}
Otherwise, the phase shift will not be accurately predicted by the 1D calculation and a 3D calculation is necessary. In the experiment, the phase shift can nevertheless be accurately measured and adjusted by slightly adjusting $\omega_\perp$.

Due to the admixture of other hyperfine levels to the qubit states, the scattering lengths $a^{00}_s$, $a^{01}_s$, and $a^{11}_s$ are mixed. The order of magnitude of the changes in the qubit scattering lengths is
\begin{equation}
\delta a_s \sim  \frac{\Omega_R^2}{2\Delta^2} (a^{11}_s-a^{01}_s) \sim \frac{\Omega_R^2}{2\Delta^2} \cdot 0.02\, a^{11}_s,
\end{equation}
for $^{87}$Rb, which typically gives $\delta a_s \sim 10^{-4}\,a^{11}_s$, which is negligible.


\section{Conclusion}

In conclusion, we have made a realistic proposal for a collisional phase gate using microwave potentials on an atom chip. The gate is implemented for a robust qubit state pair with experimentally demonstrated coherence and trap lifetimes $\tau_c \sim \tau_t \sim 1$\,s at micron-distance from the chip. We have simulated and optimized the gate dynamics for a chip layout which we specify in detail and which can be fabricated with today's technology. We found a gate fidelity of $F=0.996$ at a gate operation time of $\tau_g=1.1$\,ms, taking many error sources into account. With a total infidelity of the order of a few $10^{-3}$, our gate meets the requirements for fault-tolerant quantum computation \cite{Steane03,Knill05}. The gate fidelity of the present proposal is limited by wave packet distortion due to collisions and the dynamics of the atoms in a non-harmonic potential. We believe that these effects can be reduced by further optimization of the potential and better control of the gate dynamics, at the expense of introducing more control parameters. The ultimate limit to the fidelity will then be given by $\exp(-\tau_g/\tau_t)\sim 0.999$ for the chip layout discussed here.

While the use of microwave potentials on atom chips is within reach, a major experimental challenge remaining is the deterministic preparation of single neutral atoms in the motional ground state of chip traps with very low occupation probability of excited states. Proposals for single atom preparation have been put forward in \cite{Diener02,Mohring05}. An important prerequisite for single atom preparation is a single atom detector on the atom chip. The realization of such a detector is currently a subject of intense experimental efforts \cite{Teper06,Treutlein06,Steinmetz06}, with on-chip optical fiber cavities being a particularly promising system \cite{Horak03,Long03}.

The authors acknowledge financial support from the EC (ACQP project). P.T. is grateful for the invitations to Trento. A.N. acknowledges additional support from the EC (grant SCALA). A.N. thanks Zbigniew Idziaszek and Uffe V. Poulsen for useful discussions, and the friendly hospitality at LMU in Munich. T.C. acknowledges support from the EC (grants SCALA and QOQIP) and NSF (grant of the ITAMP at Harvard University Physics Department and Smithsonian Astrophysical Observatory).


\begin{thebibliography}{10}

\bibitem{diVincenzo00}
D.P.~DiVincenzo, Fortschr. Phys. {\bfseries 48}, 771 (2000).

\bibitem{Reichel02}
J.~Reichel, Appl. Phys. B {\bfseries 74}, 469 (2002).

\bibitem{Folman02}
R.~Folman, P.~Kr{\"u}ger, J.~Schmiedmayer, J.~Denschlag, and C.~Henkel, Adv. At. Mol. Opt. Phys. {\bfseries 48}, 263 (2002).

\bibitem{Treutlein04}
P.~Treutlein, P.~Hommelhoff, T.~Steinmetz, T.W.~H{\"a}nsch, and J.~Reichel, Phys. Rev. Lett. {\bfseries 92}, 203005 (2004).

\bibitem{Schumm05}
T.~Schumm, S.~Hofferberth, L.M.~Andersson, S.~Wildermuth, S.~Groth, I.~Bar-Joseph, J.~Schmiedmayer, and P.~Kr{\"u}ger, Nature Physics {\bfseries 1}, 57 (2005).

\bibitem{Hommelhoff05}
P.~Hommelhoff, W.~H{\"a}nsel, T.~Steinmetz, T.W.~H{\"a}nsch, and J.~Reichel, New J. Phys. {\bfseries 7}, 3 (2005).

\bibitem{Lev03}
B.~Lev, Quant. Inf. Comp. {\bfseries 3}, 450 (2003).

\bibitem{Groth04}
S.~Groth, P.~Kr\"uger, S.~Wildermuth, R.~Folman, T.~Fernholz, J.~Schmiedmayer, D.~Mahalu, and I.~Bar-Joseph, Appl. Phys. Lett. {\bfseries 85}, 2980 (2004).

\bibitem{Sorensen04}
A.S.~S{\o}rensen, C.H.~van der Wal, L.I.~Childress, and M.D.~Lukin, Phys. Rev. Lett. {\bfseries 92}, 063601 (2004).

\bibitem{Calarco00}
T.~Calarco, E.A.~Hinds, D.~Jaksch, J.~Schmiedmayer, J.I.~Cirac, and P.~Zoller, Phys. Rev. A {\bfseries 61}, 022304 (2000).

\bibitem{Negretti03}
A.~Negretti, T.~Calarco, M.A.~Cirone, and A.~Recati, Eur. Phys. J. D {\bfseries 32}, 119 (2005).

\bibitem{Steane03}
A.~Steane, Phys. Rev. A {\bfseries 68}, 42322 (2003).

\bibitem{Knill05}
E. Knill, Phys. Rev. A {\bfseries 71}, 42322 (2005).

\bibitem{Schmidt-Kaler03}
F.~Schmidt-Kaler, S.~Gulde, M.~Riebe, T.~Deuschle, A.~Kreuter, G.~Lancaster, C.~Becher, J.~Eschner, H.~H\"affner, and R.~Blatt, J. Phys. B: At. Mol. Opt. Phys. {\bfseries 36}, 623 (2003).

\bibitem{Henkel03}
C.~Henkel, P.~Kr\"uger, R.~Folman, and J.~Schmiedmayer, Appl. Phys. B {\bfseries 76}, 173 (2003).

\bibitem{Treutlein06}
P.~Treutlein, T.~Steinmetz, Y.~Colombe, B.~Lev, P.~Hommelhoff, J.~Reichel, M.~Greiner, O.~Mandel, A.~Widera, T.~Rom, I.~Bloch, and T.W.~H{\"a}nsch, Fortschr. Phys. {\bfseries 54}, 702 (2006).

\bibitem{Bloch05}
I.~Bloch, Nature Physics {\bfseries 1}, 23 (2005).

\bibitem{Agosta89}
C.C.~Agosta, I.F.~Silvera, H.T.C.~Stoof, and B.J.~Verhaar, Phys. Rev. Lett. {\bfseries 62}, 2361 (1989).

\bibitem{Spreeuw94}
R.J.C.~Spreeuw, C.~Gerz, L.S.~Goldner, W.D.~Phillips, S.L.~Rolston, C.I.~Westbrook, M.W.~Reynolds, and I.F.~Silvera, Phys. Rev. Lett. {\bfseries 72}, 3162 (1994).

\bibitem{Collin01}
R.E.~Collin, {\itshape Foundations for Microwave Engineering}, 2nd ed. (John Wiley \& Sons, Hoboken, 2001).

\bibitem{Krueger03}
P.~Kr{\"u}ger,  X.~Luo, M.W.~Klein, K.~Brugger, A.~Haase, S.~Wildermuth, S.~Groth, I.~Bar-Joseph, R.~Folman, and J.~Schmiedmayer, Phys. Rev. Lett. {\bfseries 91}, 233201 (2003).

\bibitem{Engler00}
H.~Engler, T.~Weber, M.~Mudrich, R.~Grimm, and M.~Weidem\"uller, Phys. Rev. A, {\bfseries 62}, 031402 (2000).

\bibitem{Rizzi88}
P.A.~Rizzi, {\itshape Microwave Engineering Passive Circuits}, (Prentice Hall, Englewood Cliffs, New Jersey, 1988).

\bibitem{Harber02}
D.M.~Harber, H.J.~Lewandowski, J.M.~McGuirk, and E.A.~Cornell, Phys. Rev. A {\bfseries 66}, 053616 (2002).

\bibitem{Heinrich93}
W.~Heinrich, IEEE Trans. Microwave Theory Tech. {\bfseries 41}, 45 (1993).

\bibitem{Kunze99}
M.~Kunze and W.~Heinrich, IEEE Microwave and Guided Wave Letters {\bfseries 9}, 499 (1999).

\bibitem{Olshanii98}
M.~Olshanii, Phys. Rev. Lett. {\bfseries 81}, 938 (1998).

\bibitem{Calarco01}
T.~Calarco, J.I.~Cirac, and P.~Zoller, Phys. Rev. A {\bfseries 63}, 062304 (2001).

\bibitem{Sklarz02}
S.E.~Sklarz and D.J.~Tannor, Phys. Rev. A {\bfseries 66}, 053619 (2002).

\bibitem{Calarco04}
T.~Calarco, U.~Dorner, P.S.~Julienne, C.J.~Williams, and P.~Zoller, Phys. Rev. A {\bfseries 70}, 012306 (2004).

\bibitem{Krotov96}
V.F.~Krotov, {\itshape Global Methods in Optimal Control Theory}, Monographs and Textbooks in Pure and Applied Mathematics, Vol. {\bfseries 195}, (Marcel Dekker Inc., New York, 1996).

\bibitem{Idziaszek05}
Z.~Idziaszek and T.~Calarco, Phys. Rev. A {\bfseries 71}, 050701(R) (2005).

\bibitem{Zobay01}
O.~Zobay and B.M.~Garraway, Phys. Rev. Lett. {\bfseries 86}, 1195 (2001).

\bibitem{Colombe04}
Y.~Colombe, E.~Knyazchyan, O.~Morizot, B.~Mercier, V.~Lorent, and H.~Perrin, Europhys. Lett. {\bfseries 67}, 593 (2004).

%

\bibitem{Stern90}
A.~Stern, Y.~Aharonov, and J.~Imry, Phys. Rev. A {\bfseries 41}, 3436 (1990).

\bibitem{Zhang05}
B.~Zhang, C.~Henkel, E.~Haller, S.~Wildermuth, S.~Hofferberth, P.~Kr{\"u}ger, and J.~Schmiedmayer, Eur. Phys. J. D {\bfseries 35}, 97 (2005).

\bibitem{Gentile89}
T.R.~Gentile, B.J.~Hughey, D.~Kleppner, and T.W.~Ducas, Phys. Rev. A {\bfseries 40}, 5103 (1989).

\bibitem{Idziaszek06}
Z.~Idziaszek and T.~Calarco, arXiv:quant-ph/0604205 (2006).

\bibitem{Moore03}
M.G.~Moore, T.~Bergeman, and M.~Olshanii, Les Houches School on Quantum Gases in Low Dimensions (2003), J. Phys. IV France {\bfseries 116}, 69 (2004).

\bibitem{Diener02}
R.B.~Diener, B.~Wu, M.G.~Raizen, and Q.~Niu, Phys. Rev. Lett. {\bfseries 89}, 070401 (2002).

\bibitem{Mohring05}
B.~Mohring, M.~Bienert, F.~Haug, G.~Morigi, W.P.~Schleich, and M.G.~Raizen, Phys. Rev. A {\bfseries 71}, 053601 (2005).

\bibitem{Teper06}
I.~Teper, Y.-J.~Lin, and V.~Vuleti{\'c}, arXiv:cond-mat/0603675 (2006).

\bibitem{Steinmetz06}
T.~Steinmetz, A.~Balocchi, Y.~Colombe, D.~Hunger, T.W.~H{\"a}nsch, R.J.~Warburton, and J.~Reichel, arXiv:physics/0606231 (2006).

\bibitem{Horak03}
P.~Horak, B.G.~Klappauf, A.~Haase, R.~Folman, J.~Schmiedmayer, P.~Domokos, and E.A.~Hinds, Phys. Rev. A {\bfseries 67}, 043806 (2003).

\bibitem{Long03}
R.~Long, T.~Steinmetz, P.~Hommelhoff, W.~H{\"a}nsel, T.W.~H{\"a}nsch, and J.~Reichel, Phil. Trans. R. Soc. Lond. A {\bfseries 361}, 1375 (2003).

\end{thebibliography}
\end{document}